\documentclass[12pt]{article}

\usepackage{latexsym,amsmath,amscd,amssymb,amsthm,graphics}
\usepackage{enumerate,epsfig}
\usepackage[margin=2cm]{geometry}
 \usepackage{float}
\usepackage{graphicx}
\usepackage{framed,url,multicol}
\usepackage[usenames]{color}
\usepackage{shadow,enumerate}
\usepackage{makeidx}

\usepackage[colorlinks]{hyperref}

\usepackage[all]{xy}

\def\openone{\leavevmode\hbox{\small1\kern-3.3pt\normalsize1}}

\def\b{\begin{eqnarray}}
\def\e{\end{eqnarray}}

\begin{document}
\begin{center}

{\LARGE\textbf{Integrable negative flows of the Heisenberg ferromagnet equation hierarchy  \\}} \vspace {10mm} \vspace{1mm} \noindent

{\Large \it Rossen  I. Ivanov  \footnote{E-mail:
Rossen.Ivanov@tudublin.ie}  }

\vskip 0.8cm

School of Mathematical Sciences, Technological University Dublin,

Grangegorman Lower, Dublin D07 ADY7, Ireland

\end{center}

\vskip 1.32cm

\begin{abstract}
\noindent We study the negative flows of the hierarchy of the integrable Heisenberg Ferromagnet model and their soliton solutions. The first negative flow is related to the so-called short pulse equation. We provide a framework which generates Lax pairs for the other members of the hierarchy. The Inverse scattering, based on the dressing method is illustrated with the derivation of the one-soliton solution.
\vskip.4cm


\vskip.4cm

\end{abstract}

\section{Introduction}
\label{intro}
Since the discovery of the peaked soliton (peakon) solutions of the Camassa-Holm (CH) equation \cite{CH93,CHH94} there is an enormous amount of research dedicated to integrable nonlinear equations in a non-evolutiionary  form. The interest to these equations is mainly becouse of the huge variety of the types of their solutions, especially their singular (nonclassical) solutions. Another important nonevolutionary equation is the Degasperis-Procesi (DP) equation \cite{D-P,D-H-H,D-H-H2}, which allows for peakon and shock-peakon solutions. There are thousands of research papers dedicated to the CH, DP and related equations, here we mention only a few. A review on the subject could be found in \cite{HoIv10,HoScSt2009}. The peakon, soliton and cuspon solutions are derived for example in \cite{BBS98,BBS99,HW,L-S,M1,M2,CGI,ILO}. Analytic and numerical aspects are studied e.g. in
\cite{C00,Con1,CE,CKKT07,CS00,ELY,Hen,HoIv2010} as well as in many other publications. Both CH and DP equations are in non-evolutionary form, since they contain $xxt$-derivative. On the other hand if they are written in evolutionary form, they become nonlocal. It has been also realised that both equation can be considered as ``negative'' flows of an AKNS or ZS hierarchy \cite{AKNS,ZS1,ZS2}, e.g. \cite{CIL,CI17,ILO}.
The Fokas-Lenells equation \cite{Fo,LeFo} is another example of a non-evolutionary equation from the DNLS-hierarchy (the hierarchy of the Derivative Nonlinear Schr\"odinger Equation). Integrable non-evolutionary equations with cubic nonlinearities have been found by Z. Qiao \cite{Q06,Q07,Q09}, and V. Novikov \cite{N09}, see also the remark on the peakons of
Qiao's equation in \cite{HJPW08} and the solutions in \cite{HLS09}.  Actually the  Qiao's equation 
together with the CH equation
belong to the bi-Hamiltonian
hierarchy of equations described by Fokas and Fuchssteiner \cite{FF80}. These equations also are negative flows (after change of the $x$-variable)  of some known soliton hierarchies \cite{HJPW08,IL}. This feature has motivated us to explore other examples of integrable non-evolutionary equations by looking at the negative flows of existing hierarchies of soliton equations. 

\section{Non-evolutionary integrable equations from the hierarchy of the Heisenberg Ferromagnet model: the first negative flow}

\subsection{Lax form representation and integrability}

Let us start from the following spectral problem:
\begin{equation} i\psi_x =L\psi, \qquad i\psi_t=M\psi, \label{Lax1}
\end{equation} where the Lax \cite{Lax68} operator is like the one for the Heisenberg Ferromagnet (HF) model :
\begin{equation} \label{Lax11}
L=\lambda S_0 , \qquad \text{with} \qquad
S_0 = \left(\begin{array}{cc}
\theta & m \\ -n &  -\theta
\end{array}\right)\end{equation}
and $\theta=$ const. All functions are complex-valued. The $M$-operator for the first negative flow is taken in the form:

\begin{equation}
M= \frac{1}{2\lambda}\sigma_3 + M_0+
\lambda h S_0\end{equation} for an yet undetermined scalar function(al) $h$ and $M_0;$ $\sigma_3=\text{diag}(1,-1 )$.
Then the compatibility (zero-curvature) equation \begin{equation} \label{comp}
iL_t-iM_x+[L,M]=0
\end{equation} produces the constraints $(M_0)_{11}=-(M_0)_{22}$= const, $m=-i(M_{0,x})_{12}$ and $n=-i(M_{0,x})_{21}$.  The constant diagonal part of $M_0$ can be removed by a trivial gauge transformation, so we can take it to be zero. Then by introducing new variables $$ m=i u_x , \qquad n=iv_x,$$ we obtain for the $sl(2,\mathbb{C})$ element \[M_0 = \left(\begin{array}{cc}
0 & -u \\ -v & 0
\end{array}\right).\] Assuming further that all functions belong to the Schwartz class $\mathcal{S} (\mathbb{R}) $ in $x$ for all $t\in \mathbb{R},$ the arising differential equations are \begin{eqnarray}
i(h\theta)_x + nu+mv &=&0 \\
im_t-i(hm)_x -2 \theta u  &=&0\\
in_t-i(hn)_x  -2 \theta v &=&0.
\end{eqnarray}
The first equation is a differential constraint giving
\begin{equation} h= -\frac{1}{\theta}u v \end{equation}   and finally we arrive at the following system of coupled equations
\begin{align}
u_{xt}+ \frac{1}{\theta}(v u u_x)_x +2 \theta u  &=0\\
v_{xt}+ \frac{1}{\theta}(u v v_x)_x +2 \theta v  &=0.
\end{align}
The first obvious reduction $v= \bar{u}$ gives
\begin{equation} \label{eq_1}
u_{xt}+ \frac{1}{\theta}(|u|^2 u_x)_x +2 \theta u  =0.
\end{equation}
Another possible reduction is $u=v$ giving
\begin{equation}
u_{xt}+ \frac{1}{3\theta}(u^3)_{xx} +2 \theta u  =0.
\end{equation}
The last two equations are known as the ''short pulse equation'' and clearly have solutions when $u$ is real as well \cite{ShW,Ch,BdM,Bru}. Coupled short pulse equations and other generalisations are also at the focus of recent research \cite{Feng,HNW,M3}.

\subsection{Spectral theory and soliton solutions}

The spectral theory of the HF Lax operator is in principle well known, see for example \cite{FaTa,GVY,GY1}. We are going to repeat breafly the known facts, since the negative flows have some specifics.  From  (\ref{Lax1}) with $v=\bar{u}$ (and $n=-\bar{m}$) we have
$$ S_0^2 = (\theta^2 + |m|^2) \openone. $$
Therefore, in order to write the HF spectral problem in a canonical form, we need to change the variables, according to
\begin{equation}  dy = \sqrt{\theta^2 + |m|^2} dx \label{VC} \end{equation}  and introduce a new variable $y$ as a function of $x,t.$ Moreover, \eqref{VC} suggests that $y \to \infty$ when $x \to \infty$ and vice versa. We introduce
$$ S= \frac{1}{\sqrt{\theta^2 + |m|^2}}S_0, \qquad S^2 = \openone. $$
The newly obtained the spectral problem is
\begin{equation} \label{Gauge1} i\tilde{\psi}_y= \lambda \left(\begin{array}{cc}
\frac{\theta}{\sqrt{\theta^2 + |m|^2}} & \frac{m}{\sqrt{\theta^2 + |m|^2}} \\ \frac{\bar{m}}{\sqrt{\theta^2 + |m|^2}} &  \frac{-\theta}{\sqrt{\theta^2 + |m|^2}}
\end{array}\right)\tilde{\psi}\equiv \lambda S(y) \tilde{\psi}(y, \lambda)
\end{equation}
Without loss of generality we can assume that $\theta >0$ is a real constant, then
the matrix $S$ in the spectral problem (\ref{Gauge1}) has the properties
\begin{equation} S=S^{\dagger}, \qquad S^2=\openone, \qquad \lim_{|y|\to \infty}S = \sigma _3. \label{S}\end{equation}
The spectral problem (\ref{Gauge1})then has the precise form of the one of the hierarchy of the Heisenberg ferromagnet (HF) model \cite{FaTa,GVY,GY1}.

Due to the imposed reduction on $S$, the eigenfunction $\tilde{\psi}(y, \lambda)$ are $SU(2)$ group-valued elements with the following property, whose verification is straightforward:

\begin{equation} \label{reduction}
\tilde{\psi}^{-1}(y, \lambda)=\tilde{\psi}^{\dagger}(y, \bar{\lambda})
\end{equation}
For simplicity where possible we will be omitting the variable $t$, which can be considered as a parameter.  The functions $m(x,t)$, $u(x,t)$ are assumed to belong in the space of the rapidly decreasing functions on $\mathbb{R}$ (Schwartz class) when $x\in \mathbb{R}$ for all values of $t,$ which could be considered as a fixed parameter when we study the spectral theory. Hence under the variable change \eqref{VC} the functions are Schwartz class when $y\in \mathbb{R}$ as well. Let us introduce the notation
\begin{equation} \label{Gauge2} S = \left(\begin{array}{cc}
S_3 & S_{12} \\ \bar{S}_{12} &  - S_3
\end{array}\right) ,
\end{equation}
$S$ takes values in the $su(2)$ algebra. The inverse scattering method provides a solution for the components of $S$ in terms of the so-called scattering data. In what follows we will outline briefly this  method by using the so-called dressing procedure, \cite{GVY}.
The spectral problem  (\ref{Gauge1}) is equivalent to a canonical Zakharov - Shabat (ZS) spectral problem, \cite{ZS1,ZS2} under a gauge transformation
such that $\tilde{\psi}(y, \lambda)= H(y) \varphi(y, \lambda) $ for a non-degenerate matrix $H(y)$.  Indeed, the new spectral problem is
\begin{equation} \label{ZS}
i\partial_y\varphi(y, \lambda)+(Q(y)-\lambda \sigma_3)\varphi(y, \lambda)=0, \qquad Q(y)=iH^{-1} H_y \end{equation}
 Then, of course
 \begin{equation} \label{SH}
S=H\sigma_3 H^{-1} \end{equation}  and $H$ determines $S$ completely. Moreover, due to \eqref{S}, we can consider that $\lim_{|y| \to \infty} H(y)= \openone$ and hence  $$\lim_{|y| \to \infty} Q(y)= 0,$$  $Q(y) \in su(2)$. Due to \eqref{reduction} we have $\varphi$  and $H$ in $SU(2)$ and

\begin{equation} \label{reduction1}
\varphi^{-1}(y, \lambda)=\varphi ^{\dagger}(y, \bar{\lambda}), \qquad H^{-1}=H^{\dagger}, \qquad H H^{\dagger}= \openone.
\end{equation}
 The dressing method works as follows. Starting from a trivial (or bare) solution $Q(y,t)=0$, with its associated eigenfunction $\varphi_0(y,t,\lambda)$, we may obtain an eigenfunction $\varphi(y,t,\lambda)$ corresponding to soliton solutions, via the {\it dressing factor} $g(y,t,\lambda)$, defined by the following
\begin{equation}\label{DF}
        \varphi(y,t,\lambda) = g(y,t,\lambda) \varphi_0(y,t,\lambda).
\end{equation}
The dressing factor $g(y,t,\lambda)\in SU(2)$ is singular at each discrete eigenvalue $\lambda=\lambda_n$ and is otherwise analytic for any complex $\lambda $ and moreover from \eqref{reduction1}, \eqref{DF} and  (\ref{Gauge1})
\begin{align}
&g^{-1}(y, \lambda)=g ^{\dagger}(y, \bar{\lambda}),\label{g_reduction}\\
& i g_y +Q g - \lambda[\sigma_3 , g] =0. \label{eq_g}
\end{align}
In the one-soliton case we consider a ZS dresing factor with one pole at $\lambda=\mu$:
\begin{align}
&g(y, \lambda)g ^{\dagger}(y, \bar{\lambda})=\openone,\label{g_reduction1}\\
&g(y, \lambda)=\openone + \frac{\mu - \bar{\mu}}{\lambda - \mu}P(y),\label{g_1p}
\end{align}
where $P(y)$ is a projector, $P^2=P$.  From \eqref{eq_g}--\eqref{g_1p} we obtain
\begin{align}
&P^{\dagger}=P , \qquad  Q=(\mu - \bar{\mu}) [\sigma_3, P], \\
& iP_y + QP - \mu[\sigma_3, P]=0,  \end{align} and hence the equation for $P$ is  \begin{equation}
 iP_y - \bar{\mu}\sigma_3 P +\mu P \sigma_3 -(\mu - \bar{\mu})P\sigma_3 P=0, \label{eq_P}
\end{equation}
From \eqref{eq_g} when $\lambda=0$ we have
\begin{equation}
i g_y(y,0)+Q (y)g(t,0)=0 \quad \text{or} \quad i g_y(y,0)+H^{-1} i H_y g(y,0) =0,  \quad \frac{\partial}{\partial y}(H(y)g(y,0))=0.
\end{equation}
Therefore we can choose the gauge transform
\begin{equation} \label{H}
H(y)=g^{-1}(y,0).
\end{equation}
Assuming a rank 1 projector
$$ P(y)=\frac{| F \rangle \langle \bar{F}|  }{\langle \bar{F}| F \rangle}, \qquad | F \rangle =(F_1, F_2)^T$$
one can verify that \eqref{eq_P} is satisfied if $ | F \rangle $ satisfies the {\it bare} equation ($Q(y)=0$)
$$ i\partial_y | F \rangle +(0- \bar{\mu} \sigma_3) | F \rangle =0. $$
This way we can take $ | F \rangle = \varphi_0(y, \bar{\mu})| F_0 \rangle $ where $ |F_0 \rangle =(F_{01}, F_{02})^T$ is a constant vector and
\begin{equation}\varphi_0(y, \lambda)=e^{-i\sigma_3 \left( \lambda y + \frac{t}{2\lambda} \right)} \label{phi_0} \end{equation}
is the solution of the {\it bare } spectral problem . Letting $\mu=\alpha + i \omega$ one can evaluate the quantities
\begin{align}
& \Delta =\langle \bar{F}| F \rangle=|F_1|^2+|F_2|^2=2|F_{01}| |F_{02}| \cosh \xi, \\
&\xi(y,t)=2\omega y -\frac{ \omega t}{\alpha^2 + \omega^2 } + \xi_0, \qquad \xi_0=\ln \left| \frac{F_{02}}{F_{01}}\right|=\text{const}, \\
&P=\frac{1}{\Delta} \left(\begin{array}{cc}
|F_1|^2 &\bar{F}_1 F_2 \\ \bar{F}_2 F_1 & |F_2|^2
\end{array}\right). \label{P}
\end{align}
Furthermore, $$g(y,0)=\openone + \frac{\bar{\mu}-\mu}{\mu}P$$ and from \eqref{SH}, \eqref{H} we obtain
\begin{equation}
S= \left( \openone + \frac{\mu -\bar{\mu}}{\bar{\mu}}P \right) \sigma_3 \left( \openone + \frac{\bar{\mu}-\mu}{\mu}P \right).
\end{equation}
This is the one-soliton solution, related to eigenvalues $\alpha\pm i\omega$ and dispersion law $\frac{1}{2 \lambda}\sigma_3$. For the Heisenberg ferromagnet an analogous result could be found for example in \cite{FaTa,GVY}. Explicitly the components of $S$ are
\begin{align} S_3=\frac{\theta}{\sqrt{\theta^2 + |m|^2}} =&1-\frac{2\omega^2}{(\alpha^2+\omega^2)\cosh^2 \xi(y,t) }\label{S3}\\
S_{12}=\frac{m}{ \sqrt{\theta^2+|m|^2}}=& -\frac{2i\omega e^{i \eta }}
{(\alpha^2+\omega^2)\cosh^2 \xi} \left(\omega \sinh \xi +i\alpha \cosh \xi    \right) \label{S12}\\
\eta(y,t)= &2\alpha y + \frac{\alpha t }{\alpha^2 + \omega^2} + \eta_0, \qquad \eta_0 = \arg  \frac{\bar{F}_{01}F_{02}}{|F_{01}F_{02}|}=\text{const}.  \end{align} The variable change $x=X(y,t) $ could be obtained formally as $X= \frac{1}{\theta}\int S_3(y,t) dy$ which in this case gives
 \begin{align}
x=X(y,t) =&\frac{y}{\theta}-\frac{\omega}{\theta(\alpha^2+\omega^2)}\left[\tanh \left(2\omega y-\frac{\omega t}{\alpha^2+\omega^2}+ \xi_0 \right) +1\right]. \label{var change}\end{align}
According to (\ref{S3}) the RHS has to be everywhere positive, which necessitates $|\alpha|>|\omega|$. Note that in parametric form $$ m(X(y,t),t) =\theta \frac{S_{12}}{S_3}=  -\frac{2i\theta \omega e^{i \eta(y,t) }}
{(\alpha^2+\omega^2)\cosh^2 \xi(y,t) - 2 \omega^2} \left(\omega \sinh \xi(y,t) +i\alpha \cosh \xi(y,t)    \right)   $$ is nonsingular only if $|\alpha|>|\omega|$.
The scattering data for this 1-soliton solution are provided by 4 real constants: two of them, $\alpha$ and $\omega$ related to the discrete eigenvalue $\mu = \alpha + i \omega$ and the other two, $\xi_0, \eta_0$ are related to the choice of initial position and initial phase of the soliton.
The inconvenience here is that the solution of \eqref{eq_1} is $$u(x,t)=-i\int m(x,t) dx, \qquad u(X(y,t),t) = -i \int S_{12}(y,t) dy$$ and requires an extra integration. This difficulty could be avoided as follows. The eigenfunction of \eqref{Gauge1} is
$$\tilde{\psi}(y, \lambda)=H(y) \varphi(y, \lambda) = g^{-1}(y, 0)g(y, \lambda) \varphi_0(y,\lambda) $$  where the time- dependent $\varphi_0(y,t,\lambda)$ is given in \eqref{phi_0}. We can write then
$$\tilde{\psi}(y, \lambda)e^{i \sigma_3 \frac{t}{2\lambda}}= g^{-1}(y, 0)g(y, \lambda) e^{-i\lambda \sigma_3 y}. $$ The quantity $\Psi(y, \lambda)=\tilde{\psi}(y, \lambda)e^{i \sigma_3 \frac{t}{2\lambda}} $ is still an eigenfunction of \eqref{Gauge1} and most importantly,
\begin{equation} \Psi(y, \lambda)= g^{-1}(y, 0)g(y, \lambda) e^{-i\lambda \sigma_3 y}, \label{dres2}  \end{equation} giving $$\lim_{\lambda \to 0}  \Psi(y, \lambda) = \openone . $$ Thus one can expand about $\lambda=0$:
\begin{equation} \label{raz} \Psi(y, \lambda)=  \openone + \lambda \dot{\Psi}(y,0)+\lambda^2\frac{1}{2}\ddot{\Psi}+\ldots, \end{equation}  where the dot is a notation for a derivative with respect to $\lambda$. From \eqref{dres2}
\begin{equation} \label{dotPsi} \dot{\Psi}(y,0)= -i \sigma_3 y - \frac{\mu - \bar{\mu}}{|\mu|^2}P(y) .\end{equation}
The \eqref{Gauge1} can be written also in the form
\begin{align} \label{Gauge2} i\Psi_y= \lambda \left(\begin{array}{cc}
\theta X_y(y,t) & i\partial_y u(X(y,t),t) \\ -i\partial_y \bar{u}(X(y,t),t)
&  -\theta X_y(y,t)
\end{array}\right)\Psi (y, \lambda)= \lambda  \left(\begin{array}{cc}
\theta X & i u \\ -i\bar{u}
&  -\theta X
\end{array}\right)_y\Psi
\end{align}
From \eqref{raz} we obtain
$$ \left(\begin{array}{cc}
\theta X & i u \\ -i\bar{u}
&  -\theta X
\end{array}\right)  = i \dot{\Psi}(y, 0)= \sigma_3 y -i \frac{\mu - \bar{\mu}}{|\mu|^2}P(y) $$ or after some straightforward calculations
$$ \theta X(y,t)=[ i \dot{\Psi}(y, 0)]_{11}=y-i \frac{\mu - \bar{\mu}}{|\mu|^2}[P(y)]_{11}=y+\frac{\omega e^{-\xi}}{(\alpha^2+\omega^2) \cosh \xi}. $$
Taking into account the identity $$\frac{ e^{-\xi}}{\cosh \xi} =1-\tanh \xi$$ we obtain
$$ \theta X(y,t)=y-\frac{\omega }{\alpha^2+\omega^2}\tanh \xi + \text{const} $$
which is equivalent to \eqref{var change} because $X(y,t)$ is determined only up to an overall additive constant.

Similarly, we obtain the solution in parametric form
$$ u(X(y,t),t)=[ \dot{\Psi}(y, 0)]_{12}=- \frac{\mu - \bar{\mu}}{|\mu|^2}[P(y)]_{12}=\frac{-i\omega e^{i\eta(y,t)}}{(\alpha^2+\omega^2) \cosh \xi(y,t)} .$$
Note that $$ S_{12}=i\frac{\partial}{\partial y} u(X(y,t),t)=\frac{-2i\omega e^{i\eta(y,t)}(\omega \sinh \xi + i \alpha \cosh \xi)}{(\alpha^2+\omega^2) \cosh^2 \xi(y,t)}, $$ which is exactly as in \eqref{S12}. The above representation can be viewed as a solution in parametric form, where $y$ is the parameter. From (\ref{var change}) one can see that when $\theta=0$ there are no smooth soliton solutions with these boundary conditions at $x\to \pm \infty$. The experience with the CH and DP equations suggests that there might be singular solutions when $\theta=0$.

\section{ The second negative flow}

The second negative flow has a dispersion law $\frac{1}{2\lambda^2}\sigma_3$ which is proportional to the {\it negative} second power of the spectral parameter $\lambda.$ In this subsection for convenience we use same letters for different albeit analogous quantities. We denote $m=iu_{xx},$ $n=iv_{xx},$
the $L$ operator is as defined in \eqref{Lax1}--\eqref{Lax11}, the $M$-operator is written as expansion in $\lambda$ including also $\lambda^{-1}$ and $\lambda^0$ while the $\lambda^1$ contribution is taken to be proportional to $L.$  The compatibility condition \eqref{comp} gives the expression \begin{equation}
M= \frac{1}{2\lambda^2}\sigma_3 + \frac{1}{\lambda}\left(\begin{array}{cc}
0 & -u_x \\
-v_x &  0
\end{array}\right)+ \left(\begin{array}{cc}
-u_xv_x & 2i\theta u \\
 -2i \theta v &  u_x v_x
\end{array}\right)  + \lambda h S_0.
\end{equation}
where in addition
\begin{equation} h= 2i(u v_x-v u_x) \end{equation}  and the equations
\begin{eqnarray}
i u_{xxt}+ 2\big((u v_x-v u_x)u_{xx}\big)_x +2u_xv_xu_{xx} +4 \theta^2 u  &=&0 \label{2-1} \\
-iv_{xxt} -2 \big((u v_x-v u_x)v_{xx}\big)_x +2 u_x v_x v_{xx} +4 \theta^2 v  &=&0. \label{2-2}
\end{eqnarray}
The obvious reduction $v= \bar{u}$ leads to a single equation
\begin{equation} \label{eq_2neg}
iu_{xxt}+2\big((u \bar{u}_x-\bar{u} u_x)u_{xx}\big)_x +2|u_x|^2 u_{xx} +4 \theta^2 u  =0 .
\end{equation}
One can use the dressing method results to obtain the solutions of \eqref{eq_2neg}. Using the expansion \eqref{raz} we have as before
\begin{equation} \label{Xyt}
X(y,t)=\frac{i}{\theta} \dot{\Psi}_{11}(y,t, 0)=-\frac{i}{\theta} \dot{\Psi}_{22}(y,t, 0), \quad u_x(X(y,t),t)=\dot{\Psi}_{12}, \quad v_x(X(y,t),t)=-\dot{\Psi}_{21}.
\end{equation} Moreover, the next order gives $\frac{1}{2} \ddot{\Psi}_x=\dot{\Psi}_x \dot{\Psi}$ and componentwise
$$ \frac{1}{2} \ddot{\Psi}_{12,x}=\dot{\Psi}_{11,x}\dot{\Psi}_{12}+\dot{\Psi}_{12,x}\dot{\Psi}_{22}.$$
This last expression could be integrated once by using \eqref{Xyt} as well as integration by parts, giving
\begin{equation} \label{Uyt}
u(X(y,t),t)=\frac{i}{2\theta} \left( \frac{1}{2} \ddot{\Psi}_{12}-\dot{\Psi}_{12}\dot{\Psi}_{22} \right).
\end{equation}
By taking the $21$-component, similarly we have
\begin{equation} \label{Vyt}
v(X(y,t),t)=\frac{i}{2\theta} \left( \frac{1}{2} \ddot{\Psi}_{21}-\dot{\Psi}_{21}\dot{\Psi}_{11} \right).
\end{equation} This is consistent with the reduction $u=\bar{v}$ which necessitates $\dot{\bar{\Psi}}_{11}=\dot{\Psi}_{22},$ $\dot{\bar{\Psi}}_{21}=-\dot{\Psi}_{12}$ and $\ddot{\bar{\Psi}}_{21}=-\ddot{\Psi}_{12}.$

Because of the new $M$-operator, instead of \eqref{phi_0} we have a {\it bare} solution \begin{equation}\varphi_0(y,t, \lambda)=e^{-i\sigma_3 \left( \lambda y + \frac{t}{2\lambda^2} \right)} \label{phi_0n}, \qquad |F\rangle=\varphi_0(y,t, \bar{\mu})|F_0\rangle. \end{equation} The projector $P$ again is given in terms of the components of $F$ as in \eqref{P}. The eigenfunction $\Psi$ is defined as in \eqref{dres2}, the expression \eqref{dotPsi} is the same, in addition

\begin{equation} \label{ddotPsi} \ddot{\Psi}(y,t,0)= - y^2  \openone +  \frac{2(\mu - \bar{\mu})}{\mu |\mu|^2}P(y,t)(i \mu y \sigma_3 -  \openone) .\end{equation}

We obtain the one-soliton solution of \eqref{eq_2neg} corresponding to the pair of eigenvalues $\alpha \pm i \omega$ in parametric form as
$$ u(X(y,t),t)=\frac{\omega e^{i\eta(y,t)}( \alpha \cosh \xi+i\omega \sinh \xi)}{2\theta(\alpha^2+\omega^2)^2 \cosh^2 \xi(y,t)}, $$ where

\begin{align}
\eta(y,t)= &2\alpha y + \frac{(\alpha^2 - \omega^2) t }{(\alpha^2 + \omega^2)^2} + \eta_0, \qquad \eta_0 = \arg  \frac{\bar{F}_{01}F_{02}}{|F_{01}F_{02}|}=\text{const},\\
\xi(y,t)=&2\omega y -\frac{ 2\alpha\omega t}{(\alpha^2 + \omega^2)^2 } + \xi_0, \qquad \xi_0=\ln \left| \frac{F_{02}}{F_{01}}\right|=\text{const}, \\
 x=& X(y,t)=\frac{y}{\theta}-\frac{\omega }{\theta(\alpha^2+\omega^2)}\tanh \xi + \text{const}.
\end{align}

\section{The third negative flow}

The next order negative flows could be obtained in a similar way. The third negative flow has a dispersion law $\frac{1}{2\lambda^3}\sigma_3.$ The $L$-operator is like in \eqref{Lax1}--\eqref{Lax11} however $m=iu_{xxx},$ $n=iv_{xxx},$
the $M$-operator due to \eqref{comp} is \begin{align}
M &= \frac{1}{2\lambda^3}\sigma_3 + \frac{1}{\lambda^2}\left(\begin{array}{cc}
0 & -u_{xx} \\
-v_{xx} &  0
\end{array}\right)+\frac{1}{\lambda} \left(\begin{array}{cc}
-u_{xx}v_{xx} & 2i\theta u_x \\
 -2i \theta v_x &  u_{xx} v_{xx}
\end{array}\right)  + \left(\begin{array}{cc}
0 & R[u,v] \\
 P[u,v] &  0
\end{array}\right)   \nonumber \\
& +2i\theta (u_x v_{xx}-v_x u_{xx})\sigma_3 +  \lambda h S_0.
\end{align}
The compatibility condition \eqref{comp} gives in addition
\begin{align} P&=4\theta^2 v + 2\partial_x^{-1}(u_{xx}v_{xx}v_{xxx})   \\  R&= 4\theta^2 u + 2\partial_x^{-1}(v_{xx}u_{xx}u_{xxx})    \\  h&= 4\theta(u v_{xx}+v u_{xx}-u_x v_x)-\frac{1}{\theta}u_{xx}^2 v_{xx}^2 +\frac{2}{\theta}  v_{xx}\partial_x^{-1}(v_{xx}u_{xx}u_{xxx})  +\frac{2}{\theta}  u_{xx}\partial_x^{-1}(u_{xx}v_{xx}v_{xxx})\end{align}  and the equations
\begin{eqnarray}
 u_{xxxt}-(hu_{xxx})_x -8\theta^3 u  -4\theta  \partial_x^{-1}(v_{xx}u_{xx}u_{xxx}) -4 \theta (u_x v_{xx} - v_x u_{xx})u_{xxx} &=&0\\
v_{xxxt}-(hv_{xxx})_x -8\theta^3 v  -4\theta  \partial_x^{-1}(u_{xx}v_{xx}v_{xxx}) +4 \theta (u_x v_{xx} - v_x u_{xx})v_{xxx} &=&0.
\end{eqnarray}
The obvious reduction $v= u$ leads to a single local equation
\begin{equation} \label{eq_2neg}
u_{xxxt}-\left(\big(8\theta u u_{xx}-4 \theta u_x^2 + \frac{1}{3\theta}u_{xx}^4\big)u_{xxx}\right)_x -\frac{4\theta}{3}u_{xx}^3  -8 \theta^3 u  =0
\end{equation} which contains nonlinear terms involving $u$ and its derivatives ip to the 5th power.

\section{Conclusions}

We have presented several integrable systems arising as negative flows of the well known HF hierarchy, which is gauge-equivalent to the hierarchy of the Nonlinear Schr\"odinger equation. We have outlined the application of the dressing method for the calculation of the soliton solutions. The multisoliton solutions could be obtained from the same scheme with
$$ g(y, \lambda)=\openone + \sum_k \frac{A_k (y)}{\lambda - \mu_k}  $$
i.e. factors with several poles \cite{GVY}.
The limit $\theta \to 0$ also deserves special attention since the solutions in this case are likely to leave the Schwartz class of functions. The system \eqref{2-1}-\eqref{2-2} for example when $ \theta=0$ contains only cross-coupled terms. A comparison with the (quite possibly nonintegrable) cross-coupled system of two CH equations from \cite{CHIP}, shows that one could expect interesting and unexpected properties due to the cubic-nonlinear interactions. The quadratic nonlinearities of the system in \cite{CHIP} for example lead to ``waltzing'' pairs of peakons, moving together, cf. also with the results in \cite{HHI}.
We mention also that there are interesting integrable examples with Lax representation containing both positive and negative powers of the spectral parameter like in \cite{GIS}. The HF hierarchy in a general setting, related to Lax operators in homogeneous spaces and the associated structures including the recursion operators are studied in \cite{GY1,GY2}.

\subsection*{Acknowledgements}

The author is thankful to Prof. V. Gerdjikov for many useful discussions.


\begin{thebibliography} {}\rm

\bibitem{AKNS}
M.~J.~Ablowitz, D.~J.~Kaup, A.~C.~Newell, and H.~Segur, The inverse scattering transform - Fourier analysis for nonlinear
problems, {\it Stud. Appl. Math.} {\bf 53} (1974), 249 --315.

\bibitem{BBS98} R. Beals, D. Sattinger and J. Szmigielski, Acoustic scattering and the extended Korteweg-de Vries hierarchy.  {\it Adv. Math.} {\bf 140},
190--206 (1998); arXiv:solv-int/9901007

\bibitem{BBS99}  R. Beals, D. Sattinger and J. Szmigielski,  Multi-peakons and a theorem of Stieltjes.  {\it Inv. Problems} {\bf 15}, L1-L4 (1999); arXiv:solv-int/9903011

\bibitem{BdM} A. Boutet de Monvel, D. Shepelsky and L. Zielinski, The short pulse equation by a Riemann-Hilbert approach, {\it Lett. Math. Phys.} {\bf 107} (2017) 1345--1373; arXiv:1608.02249 [nlin.SI]

\bibitem{Bru} J.C. Brunelli, The bi-Hamiltonian structure of the short pulse equation, {\it Phys. Lett. A} {\bf 353} (2006) 475--478; arXiv:nlin/0601014 [nlin.SI]


\bibitem{CH93} R. Camassa and D. Holm, An integrable shallow water equation with peaked solitons, {\it Phys. Rev. Lett.} {\bf 71} (1993) 1661--1664;  	arXiv:patt-sol/9305002

\bibitem{CHH94} R. Camassa, D. Holm and J. Hyman, A new integrable shallow water equation, {\it Adv. Appl. Mech.} {\bf 31} (1994) 1--33.

\bibitem{Ch} Y. Chung, C.K.R.T. Jones, T. Sch\"afer and C.E. Wayne, Ultra-short pulses in linear and nonlinear media, {\it Nonlinearity} {\bf 18}(3) (2005) 1351--1374.

\bibitem{C00} A. Constantin, Existence of permanent and breaking waves for a shallow water equation: a geometric approach, {\it  Ann. Inst. Fourier
(Grenoble)} {\bf 50} (2000) 321--362.

\bibitem{Con1}
A. Constantin, Finite Propagation Speed for the Camassa-Holm Equation, {\it J. Math. Phys.} {\bf  46} (2005) 023506 (4 pages).

\bibitem{CE}
A.~Constantin and J.~Escher, Wave breaking for nonlinear nonlocal shallow water equations, {\it Acta Mathematica} {\bf 181} (1998) 229--243.

\bibitem{CGI}
A.~Constantin, V.~Gerdjikov and R.~Ivanov, Inverse scattering transform for the Camassa-Holm equation, {\em Inverse Problems} {\bf 22} (2006), 2197 -- 2207;  arXiv:nlin.SI/0603019

\bibitem{CIL} A. Constantin, R. Ivanov  and J.  Lenells, Inverse scattering transform for the Degasperis-Procesi equation, {\it Nonlinearity} {\bf 23} (2010) 2559--2575; doi:10.1088/0951-7715/23/10/012; arXiv:1205.4754 [nlin.SI]

\bibitem{CI17} A. Constantin and R. Ivanov, Dressing Method for the Degasperis-Procesi Equation, {\it Stud. Appl. Math.} {\bf 138} (2017) 205--226,  DOI: 10.1111/sapm.12149, arXiv:1608.02120 [nlin.SI]

\bibitem{CKKT07} A. Constantin, T.  Kappeler, B. Kolev and P. Topalov, On geodesic exponential maps of the Virasoro group, {\it Annals of Global Analysis and Geometry} {\bf 31} (2007) 155--180.

\bibitem{CS00} A. Constantin and W. Strauss, Stability of peakons, {\it Commun. Pure Appl. Math.} {\bf 53} (2000) 603--610.

\bibitem{CHIP} C. Cotter, D. Holm, R. Ivanov and J. Percival, Waltzing peakons and compacton pairs in a cross-coupled Camassa-Holm equation, {\it J. Phys. A: Math. Theor.} {\bf 44} (2011) 265205 (28pp); arXiv:1103.3326 [nlin.CD]

\bibitem{D-P} A.~Degasperis and M.~Procesi, Asymptotic integrability, in {\it Symmetry and Perturbation Theory}, edited by A. Degasperis and G.
Gaeta, World Scientific (1999), pp. 23--37.

\bibitem{D-H-H}
A.~Degasperis, D.~Holm and A.~Hone, A new integrable equation with peakon solutions, {\it Theor. Math. Phys.} {\bf 133} (2002), 1461--1472; arXiv:nlin/0205023 [nlin.SI]

\bibitem{D-H-H2}
A.~Degasperis, D.~Holm and A.~Hone, Integrable and non-integrable equations with peakons, in {\it Nonlinear Physics: Theory and
Experiment} (eds: M. Boiti et al.) World Scientific Publishing 2007, 37 -- 43.

\bibitem{ELY}
J.~Escher, Y.~Liu and Z.~Yin, Global weak solutions and blow-up structure for the Degasperis-Procesi equation, {\it J. Funct. Anal.} {\bf 241} (2006), 457--485.

\bibitem{FaTa}
L. Faddeev and L. Takhtadjan, {\it The Hamiltonian Approach to Soliton Theory}, Springer Verlag, Berlin 1987.

\bibitem{Feng} B.-F. Feng, Complex short pulse and coupled complex short pulse equations, Physica D {\bf 297} (2015) 62--75.

\bibitem{Fo} A.S. Fokas, On a class of physically important integrable equations, {\it Physica D} {\bf 87} (1995) 145--150.

\bibitem{FF80} A. Fokas and B. Fuchssteiner, On the structure of symplectic operators and hereditary symmetries, {\it Lett. Nuovo Cimento}
\textbf{28} (1980) 299--303.

\bibitem{GIS} V.S. Gerdjikov, R.I. Ivanov and A.A. Stefanov, Riemann-Hilbert problem, integrability and reductions, {\it J. Geometric Mech.} {\bf 11} (2019) 167--185. doi: 10.3934/jgm.2019009; arXiv:1902.10276 [nlin.SI]

\bibitem{GVY}
V. Gerdjikov, G. Vilasi and A. Yanovski, {\it Integrable Hamiltonian Hierarchies. Spectral and Geometric Methods}, Lecture Notes in Physics {\bf 748} Springer Verlag, Berlin, Heidelberg, New York, 2008.

\bibitem{GY1} V.S. Gerdjikov and A.B. Yanovski, Gauge covariant theory of the generating operator. I. {\it Commun. Math. Phys.} {\bf 103} (1986) 549--568 ; DOI:10.1007/BF01211165

\bibitem{GY2} V.S. Gerdjikov and A.B. Yanovski, Gauge covariant formulation of the generating operator. 2. Systems on homogeneous spaces, {\it Phys. Lett. A} {\bf 110} (1985) 53--58.

\bibitem{Hen} D. Henry, Compactly supported solutions of the Camassa-Holm equation, {\it J. Nonlin. Math. Phys.} {\bf 12} (2005) 342--347.

\bibitem{HHI} D. Henry, D. Holm and R. Ivanov, On the persistence properties of the Cross-Coupled Camassa-Holm system,{\it  J. of Geometry and Symmetry in Physics}, {\bf 32} (2013) 1--13; arXiv:1311.2127 [math.AP]

\bibitem{HoIv10} D. Holm and R. Ivanov,  Smooth and peaked solitons of the CH equation, {\it J. Phys. A: Math. Theor.} {\bf 43} (2010) 434003 (18pp); arXiv:1003.1338 [nlin.CD]

\bibitem{HoIv2010}
D.D. Holm and  R.I. Ivanov, Multi-component generalizations of the CH equation: geometrical aspects, peakons and numerical examples.
{\it J. Phys. A: Math. Theor.} {\bf 43} (2010) 492001 (20pp), doi:10.1088/1751-8113/43/49/492001;
arXiv:1009.5368 [nlin.SI]

\bibitem{HoScSt2009} D. Holm, T. Schmah and C. Stoica {\it Geometric Mechanics and Symmetry: From Finite to Infinite Dimensions}.
Oxford: Oxford University Press, 2009.

\bibitem{HW} A.~N.~W.~Hone and J.~P.~Wang, Prolongation algebras and Hamiltonian operators for peakon equations, \emph{Inverse Problems} {\bf 19} (2003), 129--145.

\bibitem{HNW}A.N.W. Hone, V. Novikov and Jing Ping Wang, Generalizations of the short pulse equation, {\it Lett. Math. Phys.} {\bf  108} (2018) 927--947; arXiv:1612.02481 [nlin.SI]

\bibitem{HJPW08} A.N.W. Hone and Jing Ping Wang, Integrable peakon equations with cubic nonlinearity, {\it J. Phys. A: Math and
Theor.} {\bf 41} (2008) 372002 (10pp); arXiv:0805.4310 [nlin.SI]

\bibitem{HLS09} A.N.W. Hone, H. Lundmark, and J. Szmigielski, Explicit multipeakon solutions of Novikov's cubically nonlinear integrable
Camassa-Holm type equation, {\it Dynamics of Partial Differential Equations} {\bf 6} (2009) 253--289; arXiv:0903.3663 [nlin.SI]

\bibitem{IL} R. Ivanov and T. Lyons, Dark solitons of the Qiao hierarchy, {\it J. Math. Phys.} {\bf 53} (2012) 123701; arXiv:1211.4249 [nlin.SI]

\bibitem{ILO} R. Ivanov, T. Lyons, N. Orr, Camassa-Holm Cuspons, Solitons and Their Interactions via the Dressing Method, {\it J. Nonlin. Sci.} {\bf 30} Number 1, (2020) 225--260; arXiv:1908.00980 [nlin.SI]

\bibitem{Lax68} P.D. Lax, Integrals of nonlinear equations of evolution and solitary waves, {\em Comm. Pure Appl. Math.} {\bf 21} (1968) 467--490.

\bibitem{LeFo}  J. Lenells and A.S. Fokas, On a novel integrable generalization of the nonlinear Schr\"odinger equation, {\it Nonlinearity} {\bf 22} (2009) 11--27; arXiv:0812.1510 [nlin.SI]

\bibitem{L-S} H.~Lundmark and J.~Szmigielski, Multi-peakon solutions of the Degasperis-Procesi equation, {\it Inverse Problems} {\bf 19}
(2003), 1241--1245;	arXiv:nlin/0503033 [nlin.SI]

\bibitem{M1} Y.~Matsuno, The $N$-soliton solution of the Degasperis-Procesi equation, {\it Inverse Problems} {\bf 21} (2005), 2085--2101; arXiv:nlin/0511029 [nlin.SI]

\bibitem{M2} Y.~Matsuno, Multisoliton solutions of the Degasperis-Procesi equation and their peakon limit, {\it Inverse Problems} {\bf 21} (2005), 1553--1570.

\bibitem{M3} Y.~Matsuno, A novel multi-component generalization of the short pulse equation and its multisoliton solutions, {\it J. Math. Phys.} {\bf 52} (2011) 123702; arXiv:1111.1792 [nlin.SI]

\bibitem{N09} V. Novikov, Generalizations of the Camassa-Holm equation, {\it J. Phys. A: Math. Theor.} {\bf 42} (2009) 342002 (14pp); arXiv:0905.2219 [nlin.SI]

\bibitem{Q06} Z. Qiao, A new integrable equation with cuspons and $W/M$-shape-peaks solitons, {\it J. Math. Phys.} {\bf 47} (2006) 112701 (9 pp).

\bibitem{Q07} Z. Qiao, New integrable hierarchy, its parametric solutions, cuspons, one-peak solitons, and $M/W$-shape peak solitons, {\it J. Math.
Phys.} {\bf 48} (2007) 112701 (19 pp).

\bibitem{Q09} Z. Qiao and L. Liu, A new integrable equation with no smooth solitons, {\it Chaos, Solitons and Fractals} {\bf 41} (2009) 587-593.

\bibitem{ShW} T. Sch\"afer and C.E. Wayne, Propagation of ultra-short optical pulse in nonlinear media, {\it Physica D} {\bf 196} (2004) 90--105.



\bibitem{ZS1}
V.E. Zakharov and A.B. Shabat,
A scheme for integrating nonlinear evolution equations of mathematical physics by the inverse scattering problem. I,
\emph{Funkts. Anal. Prilozhen.}, \textbf{8} (1974), 43--53; English translation: \emph{ Funct. Anal. Appl.}, \textbf{8} (1974), 226--235.

\bibitem{ZS2}
V.E. Zakharov and A.B. Shabat,
Integration of nonlinear equations of mathematical physics by the method of inverse scattering II.,
\emph{Funkts. Anal. Prilozhen}, \textbf{13} (1979), 13--22; English translation: \emph{ Funct. Anal. Appl.}, \textbf{13} (1979), 166--174.


\end{thebibliography}
\end{document}